\documentclass[12pt,a4paper]{article}

\usepackage[utf8x]{inputenc}
\usepackage{amssymb}   
\usepackage{graphicx}
\usepackage{color}
\usepackage[left=2cm,right=2cm,top=2cm,bottom=2cm]{geometry}
\usepackage[numbers,sort&compress]{natbib}
\usepackage[bookmarks=true,colorlinks=true,citecolor=blue,linkcolor=blue,urlcolor=magenta]{hyperref}

\title{Does the problem of global warming exist at all? \\Insight from the temperature drift induced by inevitable colored noise}

\author{V.D.~Rusov$^1$\footnote{Corresponding author: Vitaliy D. Rusov, 
E-mail: siiis@te.net.ua}, V.P.~Smolyar$^1$, M.V.~Eingorn$^{1,2}$, \\
T.N.~Zelentsova$^1$, E.P.~Linnik$^1$,  M.E.~Beglaryan$^1$, B.~Vachev$^3$}

\date{}

\begin{document}


\maketitle

\begin{center}

$^1$\textit{Department of Theoretical and Experimental Nuclear Physics, \\Odessa National Polytechnic University, 1 Shevchenko ave., Odessa 65044, Ukraine}

\vspace{0.5cm}

$^2$\textit{CREST and NASA Research Centers, North Carolina Central University,\\ Fayetteville st. 1801, Durham, North Carolina 27707, U.S.A.}

\vspace{0.5cm}

$^3$\textit{Institute for Nuclear Research and Nuclear Energy, Tsarigradsko Chaussee Blvd. 72, Sofia 1784, Bulgaria}

\end{center}

\begin{abstract}
In the present paper we state a problem of the colored noise nonremovability on the climatic 30-year time scale, which essentially changes the angle of view on the known problem of global warming.
\end{abstract}


\section{Introduction}

It is known that the historical surface temperature data set HadCRUT provides a record of surface temperature trends and variability since 1850. The most well
established version of this data set, HadCRUT3 \cite{ref01}, has been produced benefiting from recent improvements to the sea surface temperature data set which
forms its marine component and from improvements to the station records which provide the land data. A new version of this data set, HadCRUT4 \cite{ref02}, improves
and updates the gridded land-based Climatic Research Unit temperature database and virtually does not differ from its predecessor -- HadCRUT3 \cite{ref01}.

\begin{figure}[htb]
  \begin{center}
  \includegraphics[width=\linewidth]{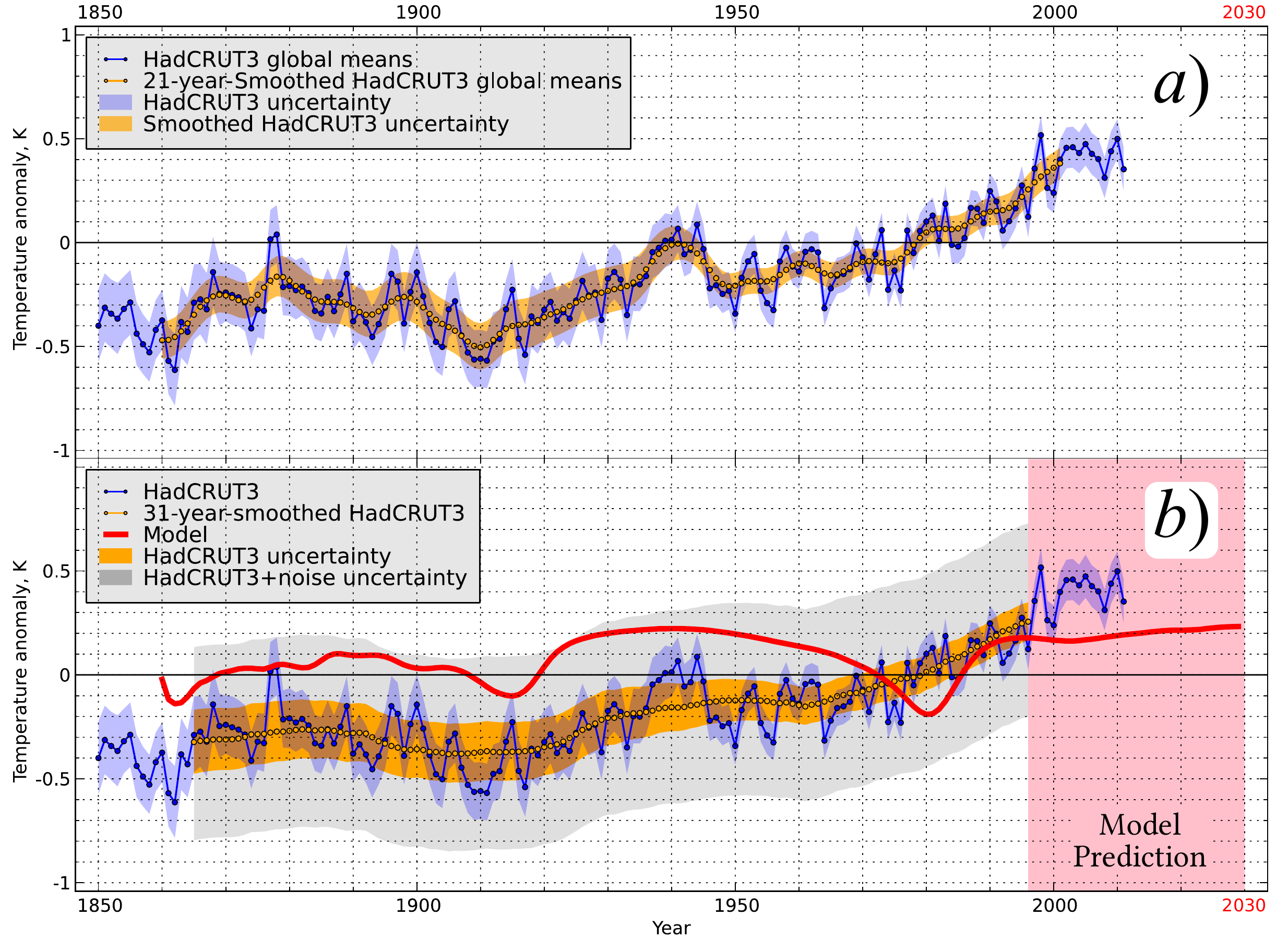}
  \end{center}
\caption{HadCRUT3 global temperature time series at annual resolution (blue) and smoothed (orange) annual resolution, obtained via averaging of the blue data
with (a) 21-year and (b) 30-year moving intervals.} \label{fig01}
\end{figure}

In the framework of both versions, basing on a careful analysis, a comprehensive set of uncertainty estimates has been derived to accompany the data and the
following perfectly clear conclusion has been made: "...Since the mid twentieth century the uncertainties in global and hemispheric mean temperatures are
small, and the temperature increase greatly exceeds its uncertainty. In earlier periods the uncertainties are larger, \emph{but the temperature increase over
the twentieth century is still \underline{significantly larger than its uncertainty}}" (Fig. \ref{fig01}a).

At the same time, analyzing the methodology of the temperature data reconstruction given in these rigorous researches \cite{ref01,ref02}, it is necessary to
note one important, in our opinion, key point about these papers. It consists in the complete absence of the commentaries on the physical reasons of ignoring
the nature and type of the noise accompanying the procedure of successive temperature averaging on different time scales. This is a major point, since every
time scale the temperature of the global climatic system is being averaged on, is characterized, in general case, by its own type of the observed noise, which
is known to be determined by the spectral density of the temperature fluctuations in the studied record. In other words, switching from one time scale
(monthly) to another (annual or thirty-year) requires the knowledge of the corresponding spectral density of the temperature fluctuations in order to estimate
the most important quantity -- the temperature variance. Meanwhile, the averaged temperature variance on the given time scale may be characterized \emph{in
certain cases} by a \emph{nonremovable} variance of the colored noise.

The effect of colored noise variance nonremovability on a given time scale means just that the colored noise is generated by the temperature fluctuations in
the climatic system. Therefore, its variance is actually a temperature variance equivalent on this time scale. Alternatively speaking, a nonremovable noise is
not an "interference" masking the real temperature, but rather a characteristic measure of the temperature fluctuations on a given time scale, which cannot be
disregarded or "cleaned out" by some kind of time-averaging procedure.

The analysis of conditions and consequences of such effect of the colored noise nonremovability typical to the power spectrum of the earth climatic system
(ECS) temperature fluctuations around tricennial time scale (the climatic state \footnote{Climate is commonly defined as a statistical ensemble of ECS states
characterized by a corresponding set of thermodynamical parameters (temperature, pressure, etc.) averaged over a long period of time. The classical period is
30 years, as defined by the World Meteorological Organization. Hence, it follows that the weather deviation from the climatic norm cannot be considered as a
climate change.}) is the primary goal of this Letter.

\section{Allan variance and a temperature drift induced by colored noise}

In order to retrieve the "true" variance magnitude for the colored noise of $1/f^{\alpha}$ type (where $\alpha \geqslant 1$) which dominates in ECS in the
climate state, it is necessary to find a good equivalent of the "original" virtual temperature record in some way. Such equivalent record must characterize the
mean global trend while not being "cleaned" by any special time-averaging procedures such as the ones used for HadCRUT3 database \cite{ref01,ref02}.

For this purpose we calculated the Allan variance \cite{ref03,ref04,ref05,ref06} for every historical temperature record observed at 5113 meteorological
stations over the world as used in the HadCRUT3 calculations

\begin{equation}
\sigma_{A}^2 (\tau) = \frac{1}{2} \left \langle \left( y_{i+1}(\tau) - y_i (\tau) \right)^2 \right \rangle,
\label{eq01}
\end{equation}

\noindent which is a variance of the first differences averaged over the time interval $\tau$ of the signal $y$. The Allan variance behavior depends on the form
of the noise power spectral density (PSD). For a noise with PSD of the $1/f^{\alpha}$ form the Allan variance (\ref{eq01}) is proportional to $\tau ^{\alpha - 1}$,
where $\alpha = 0$ stands for white (uncorrelated) noise, $\alpha = 1$ for "$1/f$" (flicker) noise, and $\alpha \geqslant 2$ for correlated low frequency
(drift) noise \cite{ref03,ref04,ref05,ref06}.

\begin{figure}[htb!]
  \begin{center}
    \includegraphics[width=12cm]{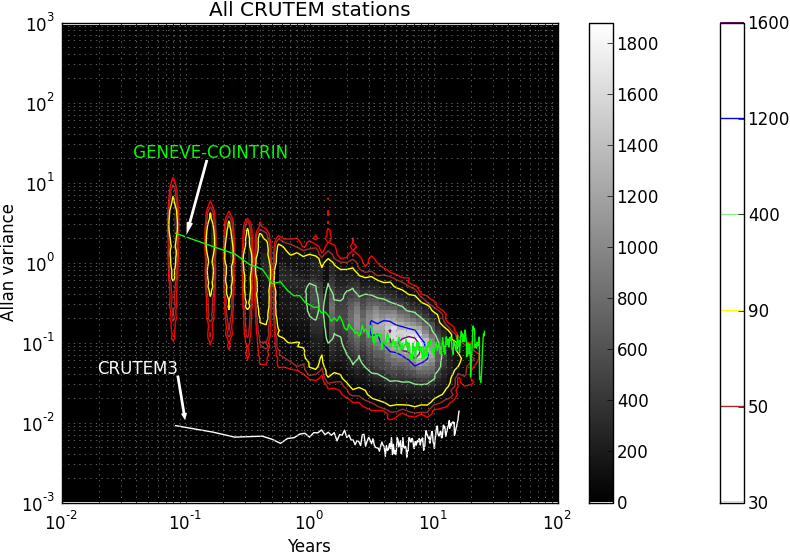}
  \end{center}
\caption{Allan variance plot for a set of temperature time series, obtained at the 5113 meteorological stations over the world. The green line corresponds to a
temperature record at Geneva-Cointrin (Switzerland) weather station. The color lines reproduce the isoclines, i.e. the lines of equal density of points that
form the general "relief" of the processed weather stations data.} \label{fig02}
\end{figure}

Analysis of the Fig. \ref{fig02} demonstrates that the experimental temperature record obtained at Geneva-Cointrin weather station over the years 1753 -- 2011
is a sufficient equivalent of the "original" virtual temperature record for our purposes, since it matches the mean global trend well and is not "cleaned" by
any kind of time-averaging procedures used in HadCRUT3 \cite{ref01,ref02}.

\begin{figure}[htb!]
  \begin{center}
    \includegraphics[width=12cm]{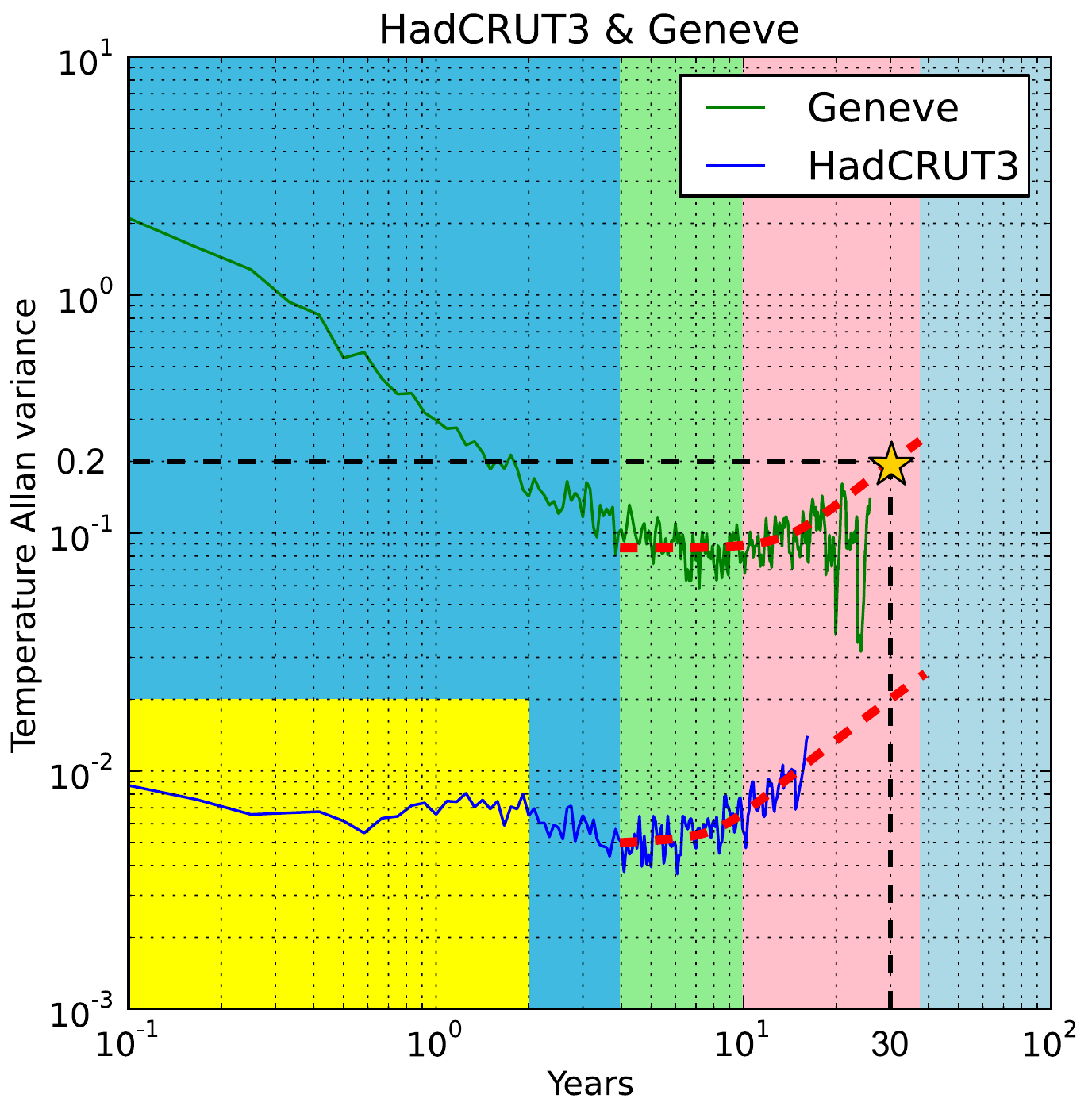}
  \end{center}
\caption{Allan variance plot for temperature records, obtained at Geneva-Cointrin weather station (the green line) and within the framework of HadCRUT3 (the blue
line).} \label{fig03}
\end{figure}

\begin{figure}[htb!]
  \begin{center}
    \includegraphics[width=12cm]{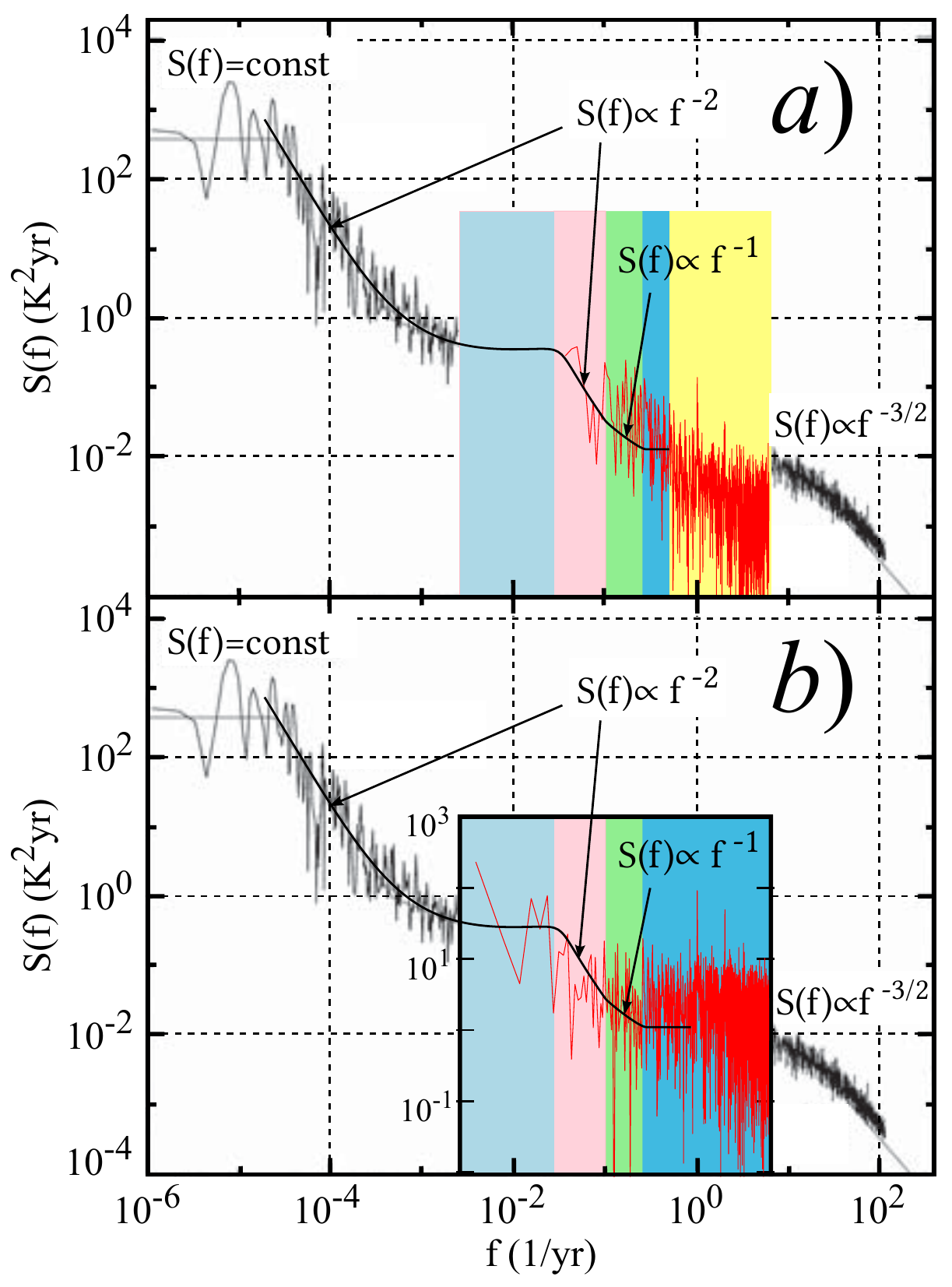}
  \end{center}
\caption{PSD of temperature fluctuations for (a) HadCRUT3 and (b) Geneva-Cointrin global temperature time series. The insets are laid over the PSD of
temperature fluctuations adopted from \cite{ref07}.} \label{fig04}
\end{figure}

Basing on the comparative analysis of Geneva-Cointrin and HadCRUT3 global temperature time series, performed by means of Allan variance (Fig. \ref{fig03}) and PSD
(Fig. \ref{fig04}), one may conclude the following. It is obvious, for example, that the time-averaging procedures used in HadCRUT3 analysis  suppressed the
Allan variance and PSD of the temperature fluctuations completely (relative to  the original Geneva-Cointrin data). In addition to that they changed the noise type, i.e. the white noise to the flicker noise (see the yellow area on Fig. \ref{fig03} and Fig. \ref{fig04}a), while the original flicker ($1/f$) and drift
($1/f^2$) noises survived, but were highly depressed (see the green and ping areas on Fig. \ref{fig03} and Fig. \ref{fig04}b). On the other hand, the original drift
noise variance apparently reaches the value of $\sigma_{A}^2$ (see the star at Fig. \ref{fig03}) around the tricennial time scale (the climatic state).

The latter means that if the drift noise around the tricennial time scale (see the star at Fig. \ref{fig03}) is nonremovable when ECS is in a state of
\emph{climate}, the standard deviation $\sigma_A$ reaches the value of $\sim 0.45$ and definitely should be taken into account when constructing the HadCRUT3
global temperature time series at smoothed annual resolution (the yellow line on Fig. \ref{fig01}a). Its adoption obviously may change the interpretation of
HadCRUT3 analysis drastically (see Fig. \ref{fig01}b, the gray color).

Considering the extreme importance of such conclusion, let us discuss the possible physical reasons of the colored noise nonremovability in climatic states
description below. For this purpose let us examine a point system, e.g., "water -- vapor", with two phase transitions taking place in it, having the interacting
order parameters $X$ and $Y$. At a point of phase transition lines intersection a potential of such system may be written down in the form of
expansion \cite{ref08,ref08a,ref08b}:

\begin{equation}
   \Phi = \Phi _0 - \alpha _1 X^2 - \alpha_2 Y^2 - \alpha_{12} XY + \beta _1 X^4 + \beta _2 Y^4 + \beta _{12} X^2 Y^2.
\label{eq02}
\end{equation}

Introducing the fluctuating forces
in the form of additive terms $\Gamma _1 (t)$ and $\Gamma _2 (t)$, where $\Gamma _1 (t)$ and $\Gamma _2 (t)$ are the Gaussian delta-correlated noises, it is
possible to pass to a system of coupled Langevin equations, following \cite{ref08,ref08a,ref08b}:

\begin{equation}
\partial X / \partial t =  - 2 \beta _{12} XY ^2 - 4 \beta _1 X^3 + 2 \alpha _1 X + \alpha _{12} Y + \Gamma_1 (t),
\label{eq03}
\end{equation}

\begin{equation}
\partial Y / \partial t =  - 2 \beta _{12} YX ^2 - 4 \beta _2 Y^3 + 2 \alpha _2 Y + \alpha _{12} X + \Gamma_2 (t).
\label{eq04}
\end{equation}

A plan for solution of such system is rather simple. The system (3)-(4) is solved numerically using, for example, the Euler method with different parameters.
The obtained numerical solutions $X(t)$ and $Y(t)$ then undergo a Fast Fourier Transform and result in a spectral density of fluctuations. In a simplest case,
when the solutions have divergent spectral characteristics, the parameters of the system, according to \cite{ref08,ref08a,ref08b}, are $\beta _{12} = 1/2$,
$\beta_1 = \alpha_{12} =1$, $\alpha_1 = \alpha_2 = \beta_2 = 0$, and the system (3)-(4) takes on the following form:

\begin{equation}
\partial X / \partial t = - XY^2 - 4X^3 + Y + \Gamma_1 (t),
\label{eq05}
\end{equation}

\begin{equation}
\partial Y / \partial t = - YX^2 + X + \Gamma_2 (t).
\label{eq06}
\end{equation}

In the absence of the external noise the solution has asymptotics $X(t) \to t^{-1/2}$ and $Y(t) \to t ^{1/2}$, when $t \to \infty$. The results of numerical
modeling presented in \cite{ref08,ref08a,ref08b} show that the frequency dependence of the spectral density $S_X (f)$ of fluctuations $X(t)$ has the form of
$1/f^{\alpha}$, where $\alpha \approx 1$. In other words, the system (\ref{eq03})-(\ref{eq04}) generates the $1/f$-noise, i.e. the flicker noise.
Meanwhile, the spectral density of the order parameter $Y(t)$ fluctuations depends on the frequency like $S_Y (f) \propto 1/f^{\mu}$, where $1.5 \leqslant \mu
\leqslant 2$, i.e. generates the drift noise.

The potential in which the system performs a random walk described by the system (\ref{eq05})-(\ref{eq06}) has the form

\begin{equation}
\Phi = \Phi_0 - XY + X^4 + ( 1 / 2 ) X^2 Y^2.
\label{eq07}
\end{equation}

As follows from (\ref{eq07}), this is a double-well potential or, more precisely speaking, a two-valley potential, i.e. a potential surface has two valleys. In
the absence of the external noise the phase trajectory of a system, depending on the initial conditions, is enclosed entirely in one of the valleys. In the
presence of an external small-amplitude noise the system performs a random walk inside the valley. Increasing the noise intensity leads to the system jumps
from one valley to the other, and $X(t)$ displays the $1/f$-noise, while the order parameter $Y(t)$ displays a $1/f^2$-noise.

Thereby this example indicates that the origin of the heat pulsations with a spectral density inversely proportional to a frequency may be related to the
interaction between the nonequilibrium phase transitions in the system of two coupled Langevin equations which transforms the white noise into two oscillation
modes with spectral densities proportional to $1/f$ and $1/f^2$ \cite{ref08,ref08a,ref08b}.

The intersection and interaction of two phase transitions is a rather widespread phenomenon. Therefore, the model (\ref{eq03})-(\ref{eq04}) \cite{ref05}
may be universal enough to serve as a basis for the explanation of the nonremovable colored noise formation mechanism in a wide range of processes involving
the nonequilibrium phase transitions. Let us show this for a point nonlinear system "water -- vapor" in the particular climatic model below.

\section{Climatic two-well  potential and bifurcation model of the Earth global climate}

In our earlier papers we proposed a bifurcation model for the energy balance of the Earth global climate. Its theoretical solution was in a good agreement with
the known experimental data on Earth surface paleotemperature evolution for the last 420 thousand years and 740 thousand years, obtained within the antarctic
projects Vostok and EPICA Dome C respectively.  In the framework of the proposed model a concept of climatic sensitivity was also introduced, which manifests a
property of temperature instability in the form of the so-called hysteresis loop, as shown in \cite{ref09,ref09a}. Basing on this concept and using the
bifurcation equation of the model, we reconstructed a time series for the global ice volume for the last 1 million years which fits well the experimental time
series of $\delta ^{18} O$ concentrations in marine sediments. A base for this result lies in a fact that an effective mechanism of climatic "cold-warm"
oscillations formation had been built into the model and was provided by the interaction of the nonequilibrium phase transitions in a "fresh water -- water
vapor" system in the Earth boundary layer \cite{ref09,ref09a}.

If we add to the stated above that our bifurcation model of the Earth global climate, developed in \cite{ref08,ref08a,ref08b}, was built upon a climatic
double-well potential

\begin{equation}
U (T, t) = \frac{1}{4} T^4 + \frac{1}{2} a(t) T^2 + b(t) T,
\label{eq08}
\end{equation}

\noindent where

\begin{equation}
a(t) = - \frac{1}{4 \delta \sigma} a_{\mu} H_{\oplus} (t),
\label{eq09}
\end{equation}

\begin{equation}
b(t) = - \frac{1}{4 \delta \sigma} \left[ \frac{\eta_{\alpha} S_0}{4} + \frac{1}{2} \beta + \frac{1}{2} b_{\mu} H_{\oplus} (t) \right],
\label{eq10}
\end{equation}

\noindent which is qualitatively identical (given $Y(t) \sim \mathrm{const}$) to a two-valley potential (\ref{eq07}), it becomes clear that ECS is really capable of
generating the colored noise which is not actually a masking ''interference'', but rather an internal property of the ECS temperature pulsations. Here
$H_{\oplus}$ is the Y-component \footnote{The physical sense of the Y-component of the Earth magnetic field adoption is discussed in \cite{ref10} in detail.} of
the Earth magnetic field intensity; $T$ is the mean global temperature of the Earth surface at the time $t$; $S_0 = 1366.2 ~ W / m^2$ is the solar constant;
$\delta = 0.95$ is the emissivity of the Earth surface; $\sigma = 5.67 \cdot 10^{-8} ~ W / (m^2 K^4)$ is the Stefan-Boltzmann constant; $\eta _{\alpha} = 0.01513
~K^{-1}$, $a_{\mu}=0.5398~W / (m^2 K^2)$, $b_{\mu}=-310~W / (m^2 K)$; finally, $\beta = 0.006 ~W / (m^2 K)$ is the carbon dioxide accumulation rate in the atmosphere normalized by a unit temperature.

Following this conclusion, let us consider a basic equation of the bifurcation model of the Earth global climate for the average Earth surface temperature on the
30-year time scale \cite{ref10,ref09,ref09a}:

\begin{equation}
\frac{m^{*}}{4 \delta \sigma}\frac{dT}{dt} = T^3 + a(t)T + b(t) + \xi(t),
\label{eq11}
\end{equation}

\noindent where the numerical value $0.129$ was used for $m^{*}$, and

\begin{equation}
\langle \xi (t) \rangle = 0, ~~~ \langle \xi (t) \xi (t') \rangle = 2 D \delta (t - t').
\label{eq12}
\end{equation}

Obviously, the solution of the stochastic differential equation (\ref{eq11}) fits into  the total uncertainty limits of HadCRUT3 global temperature time
series at smoothed annual resolution completely (Fig. \ref{fig01}b, the red line), where a nonremovable noise variance predominates. At the same time, as the
numerical integration of Eq. (\ref{eq01})
using the Runge-Kutta fourth order method with different initial conditions reveals, the obtained solution (Fig. \ref{fig01}b, the red line) is strongly stable.
It means that in the presence of a nonremovable colored noise on the climatic 30-year time scale the interpretation of the HadCRUT3 analysis data changes
drastically, replacing the known "global warming" paradigm with an alternative theory of climate as a highly stable ECS state which is characterized by a
temperature $\sim 287~K$ within the colored noise variance.

\section{Conclusion}

One of the major results of the present paper is a statement of the problem of the possible nonremovability of the colored noise on the climatic 30-year time scale,
which changes the angle of view on the known problem of global warming radically.

One of the basic climate-generating nonlinear "fresh water -- water vapor" subsystem was involved to explain the mechanism of the nonremovable colored noise
formation. The appearance of the nonremovable thermal pulsations with the spectral density inversely proportional to the frequency is explained by the interaction of
the nonequilibrium phase transitions in such system. In other words, is has been shown that ECS is really capable of generating the nonremovable colored noise
which is an internal property of the temperature pulsations and not just a masking "interference".

However it must be admitted that all the arguments adduced here apply not to a direct, but to an indirect proof of the colored noise nonremovability on the
climatic 30-year time scale. This is also true for our model representations, which, although have passed a reliable verification by fitting the known
experimental paleotemperature data on a large time scale, have the same arguments of indirect action. At the same time it should be noted that the problem of the
possible noise nonremovability formulated in the present paper makes any claims about global warming inappropriate until the problem of the colored noise
origin on the climatic 30-year time scale is definitely solved.

\section*{Acknowledgements}
This work is partially supported by EU FP7 Marie Curie Actions, SP3-People, IRSES project BlackSeaHazNet (PIRSES-GA-2009-246874).

The work of M. Eingorn was supported by NSF CREST award HRD-0833184 and NASA grant NNX09AV07A.

\bibliographystyle{plainnat}
\bibliography{Rusov-GlobalWarming}

\end{document}